\documentclass[aps,prl,reprint,amsmath,amssymb,superscriptaddress,showpacs]{revtex4-2}
\usepackage{graphicx}
\usepackage{dcolumn}
\usepackage{bm}
\usepackage[colorlinks,linkcolor=blue,anchorcolor=blue,citecolor=blue,urlcolor=blue,filecolor=blue,menucolor=blue,runcolor=blue]{hyperref}

\begin{document}
\title{Fermi Surface Nesting with Heavy Quasiparticles in the Locally Noncentrosymmetric Superconductor CeRh$_{2}$As$_{2}$}
\author{Yi Wu}
\thanks{These authors contributed equally to this paper}
\affiliation{Center for Correlated Matter and School of Physics, Zhejiang University, Hangzhou 310058, China}
\author{Yongjun Zhang}
\thanks{These authors contributed equally to this paper}
\affiliation{Institute for Advanced Materials, Hubei Normal University, Huangshi 435002, China}
\author{Sailong Ju}
\affiliation{Swiss Light Source, Paul Scherrer Institut, CH-5232 Villigen PSI, Switzerland}
\author{Yong Hu}
\affiliation{Swiss Light Source, Paul Scherrer Institut, CH-5232 Villigen PSI, Switzerland}
\author{Yanen Huang}
\affiliation{Center for Correlated Matter and School of Physics, Zhejiang University, Hangzhou 310058, China}
\author{Yanan Zhang}
\affiliation{Center for Correlated Matter and School of Physics, Zhejiang University, Hangzhou 310058, China}
\author{Huali Zhang}
\affiliation{Center for Correlated Matter and School of Physics, Zhejiang University, Hangzhou 310058, China}
\author{Hao Zheng}
\affiliation{Center for Correlated Matter and School of Physics, Zhejiang University, Hangzhou 310058, China}
\author{Guowei Yang}
\affiliation{Center for Correlated Matter and School of Physics, Zhejiang University, Hangzhou 310058, China}
\author{Evrard-Ouicem Eljaouhari}
\affiliation{Institut for Mathematische Physik, 38106 Braunschweig, Germany}
\author{Baopeng Song}
\affiliation{Institute for Advanced Materials, Hubei Normal University, Huangshi 435002, China}
\author{Nicholas C. Plumb}
\affiliation{Swiss Light Source, Paul Scherrer Institut, CH-5232 Villigen PSI, Switzerland}
\author{Frank Steglich}
\affiliation{Center for Correlated Matter and School of Physics, Zhejiang University, Hangzhou 310058, China}
\affiliation{Max Planck Institute for Chemical Physics of Solids, 01187 Dresden, Germany}
\author{Ming Shi}
\affiliation{Center for Correlated Matter and School of Physics, Zhejiang University, Hangzhou 310058, China}
\affiliation{Swiss Light Source, Paul Scherrer Institut, CH-5232 Villigen PSI, Switzerland}
\author{Gertrud Zwicknagl}
\affiliation{Institut for Mathematische Physik, 38106 Braunschweig, Germany}
\affiliation{Max Planck Institute for Chemical Physics of Solids, 01187 Dresden, Germany}
\author{Chao Cao}
\email {ccao@zju.edu.cn}
\affiliation{Center for Correlated Matter and School of Physics, Zhejiang University, Hangzhou 310058, China}\author{Huiqiu Yuan}
\email {hqyuan@zju.edu.cn}
\affiliation{Center for Correlated Matter and School of Physics, Zhejiang University, Hangzhou 310058, China}
\affiliation{Collaborative Innovation Center of Advanced Microstructures, Nanjing University, Nanjing 210093, China}
\affiliation{State Key Laboratory of Silicon Materials, Zhejiang University, Hangzhou 310058, China}
\author{Yang Liu}%
\email {yangliuphys@zju.edu.cn}
\affiliation{Center for Correlated Matter and School of Physics, Zhejiang University, Hangzhou 310058, China}
\affiliation{Collaborative Innovation Center of Advanced Microstructures, Nanjing University, Nanjing 210093, China}
\date{\today}%
\addcontentsline{toc}{chapter}{Abstract}

\begin{abstract}
 The locally noncentrosymmetric heavy fermion superconductor CeRh$_{2}$As$_{2}$ has attracted considerable interests due to its rich superconducting phases, accompanied by a quadrupole density wave and pronounced antiferromagnetic excitations. To understand the underlying physics, we here report measurements from high-resolution angle-resolved photoemission. Our results reveal fine splittings of the conduction bands related to the locally noncentrosymmetric structure, as well as a quasi-two-dimensional Fermi surface (FS) with strong $4f$ contributions. The FS exhibits nesting with an in-plane vector ($\pi/a$, $\pi/a$), which is facilitated by the van Hove singularity near $\bar{X}$ that arises from the characteristic conduction-$f$ hybridization. The FS nesting provides a natural explanation for the observed antiferromagnetic excitations at ($\pi/a$, $\pi/a$), which could be intimately connected to its unconventional superconductivity. Our experimental results are well supported by density functional theory plus dynamical mean field theory calculations, which can capture the strong correlation effects. Our study not only provides spectroscopic proof of the key factors underlying the field-induced superconducting transition, but also uncovers the critical role of FS nesting and lattice Kondo effect in the intertwined spin and charge fluctuations.
\end{abstract}

\maketitle

\par The recent discovery of heavy fermion (HF) superconductivity ($T_{c}$ $\sim$ 0.26 K) in CeRh$_{2}$As$_{2}$ has attracted broad interest \cite{Khim2021science}. The large specific jump at $T_{c}$ indicates that the Cooper pairs are formed out of the heavy quasiparticles as a result of the Kondo effect \cite{Khim2021science}. A remarkable field-induced first-order transition within the superconducting phase can be observed for a magnetic field applied along the $c$ axis \cite{Khim2021science}, which was attributed to an even-odd parity transition as a result of the local noncentrosymmetricity at the Ce sites \cite{Khim2021science,PhysRevX.AngledependenceHc2,PhysRevB2012PDW,PhysRevResearch.3.L032071,PhysRevResearch.3.023204,PhysRevResearch.3.033133,PhysRevB.105.L020505,PhysRevResearch.3.023179,PhysRevB.104.134517,arXiv2022theoryofLISB}. Such a parity transition is extremely rare for superconductors. A small interlayer hopping and larger Rashba-like spin-orbit coupling (SOC, as a result of the local noncentrosymmetricity) are thought to be essential for the parity transition of superconductivity \cite{Khim2021science,PhysRevB2012PDW,PhysRevResearch.3.L032071,PhysRevResearch.3.023204,PhysRevResearch.3.033133,PhysRevB.105.L020505,Siddiquee2022tuning}. It is therefore imperative to unravel the electronic structure underlying the peculiar superconductivity, particularly the influences of the interlayer coupling and the locally noncentrosymmetric structure.

\par Another key question in CeRh$_{2}$As$_{2}$ is the origin of the observed antiferromagnetic (AFM) fluctuations \cite{Chen2023arXiv} or order \cite{PhysRevLettAFMorder}, as well as the non-magnetic transition at $T_{0}$ $\sim$ 0.4 K \cite{Khim2021science,PhysRevX.12.011023}. Pronounced AFM excitations observed from inelastic neutron scattering (INS) \cite{Chen2023arXiv} and possible AFM order below $T_{c}$ inferred from nuclear quadrupole resonance \cite{PhysRevLettAFMorder} point towards a magnetically driven superconductivity in CeRh$_{2}$As$_{2}$, similar to other prototypical HF superconductors, such as CeCu$_{2}$Si$_{2}$ \cite{Steglich1979Superconductivity,Smidman2023RMP} and CeCoIn$_{5}$ \cite{Petrovic2001}. Interestingly, quasi-two-dimensional (2D) magnetic excitations were observed at $q_0$ = ($\pi/a$, $\pi/a$) in INS \cite{Chen2023arXiv}, which is often related to the nesting of Fermi surface (FS). On the other hand, the transition at $T_{0}$ is proposed to be an unconventional quadrupole density wave (QDW) based on renormalized band calculations \cite{PhysRevX.12.011023,Zwicknagl2016RPP}, where the nesting of the FS (involving itinerant $4f$ bands) is also necessary. Nevertheless, how the itinerant heavy quasiparticles develop and can account for the coexisting spin and charge fluctuations in CeRh$_{2}$As$_{2}$ remains an open question \cite{PhysRevB.106.L100504,PC2022HundKond,arXiv2022Violation}.

\par Here we report measurements of the quasiparticle dispersion and FS in CeRh$_{2}$As$_{2}$, from angle-resolved photoemission spectroscopy (ARPES). Although there have been a few theoretical studies on the electronic structure \cite{PhysRevResearch.3.L032071,PhysRevB.105.L020505,Ishizuka2023arXiv,PhysRevX.12.011023,PhysRevB.104.L041109}, high-resolution ARPES measurement is still lacking. Our experimental results are well supported by state-of-art calculations from density functional theory (DFT) plus dynamical mean-field theory (DMFT), and such a combination allows us to develop a thorough understanding of the relevant physics. Details of crystal growth, characterizations, ARPES measurements and calculations can be found in \cite{supplementary} (see also references \cite{Ernst2011Emerging,Wirth2016Exploring,PhysRevB.Optical,PhysRevB.47.558,PhysRevB.59.1758,PhysRevLett.77.3865,MOSTOFI2008685,Zhi2022symmetry,Haule2010Dynamical,Schwarz2002Electronic,Haule2007Quantum} therein).

\begin{figure}[ht]
\includegraphics[width=1.0\linewidth]{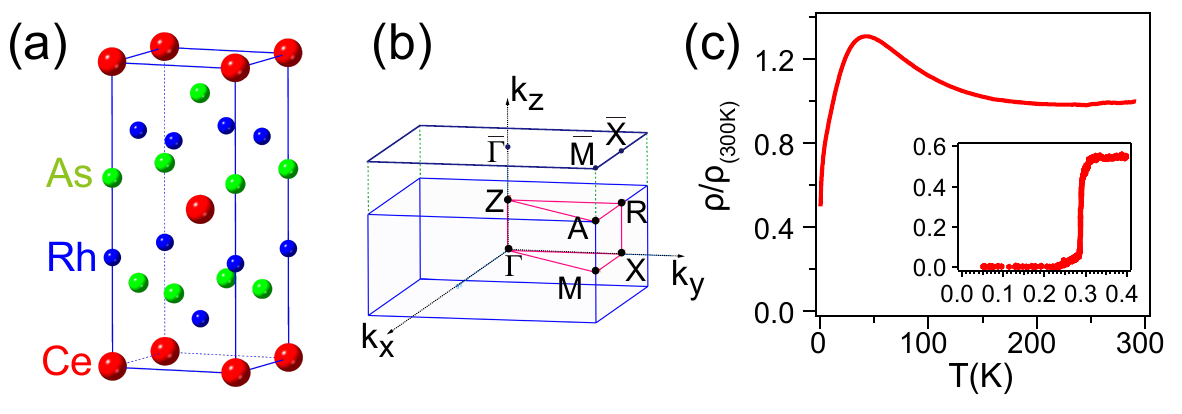}
\caption{(a) Crystal structure of CeRh$_{2}$As$_{2}$. (b) The bulk (bottom) and projected surface (top) BZ's. (c) Resistivity vs temperature. Inset is a zoomed-in view near $T_{c}$.
}
\label{fig1}
\end{figure}

\par CeRh$_{2}$As$_{2}$ crystallizes in a CaBe$_{2}$Ge$_{2}$-type tetragonal structure (see Fig.~\ref{fig1}(a)): each Ce layer is sandwiched by alternating Rh-As-Rh and As-Rh-As trilayers, which break the local inversion symmetry at the Ce sites, although the global inversion center still exists. The bulk and surface Brillouin Zone's (BZ's) are displayed in Fig.~\ref{fig1}(b). The sample quality is confirmed by the temperature-dependent resistivity (Fig.~\ref{fig1}(c)), which exhibits a broad maximum at $T^*$ $\sim$ 40 K and a superconducting transition at $T_{c}$ $\sim$ 0.24 K (zero resistivity). Note that the cleaved surfaces for ARPES measurements most likely consist of small domains with different layer terminations, which are smaller than the beam spot (Fig. S2 in \cite{supplementary}). While the mixed surface terminations lead to considerable spectral broadening, they also indicate that the strong features observed experimentally should be bulk states, as verified by comparison with calculations (see below).

\begin{figure}[ht]
\includegraphics[width=1.0\linewidth]{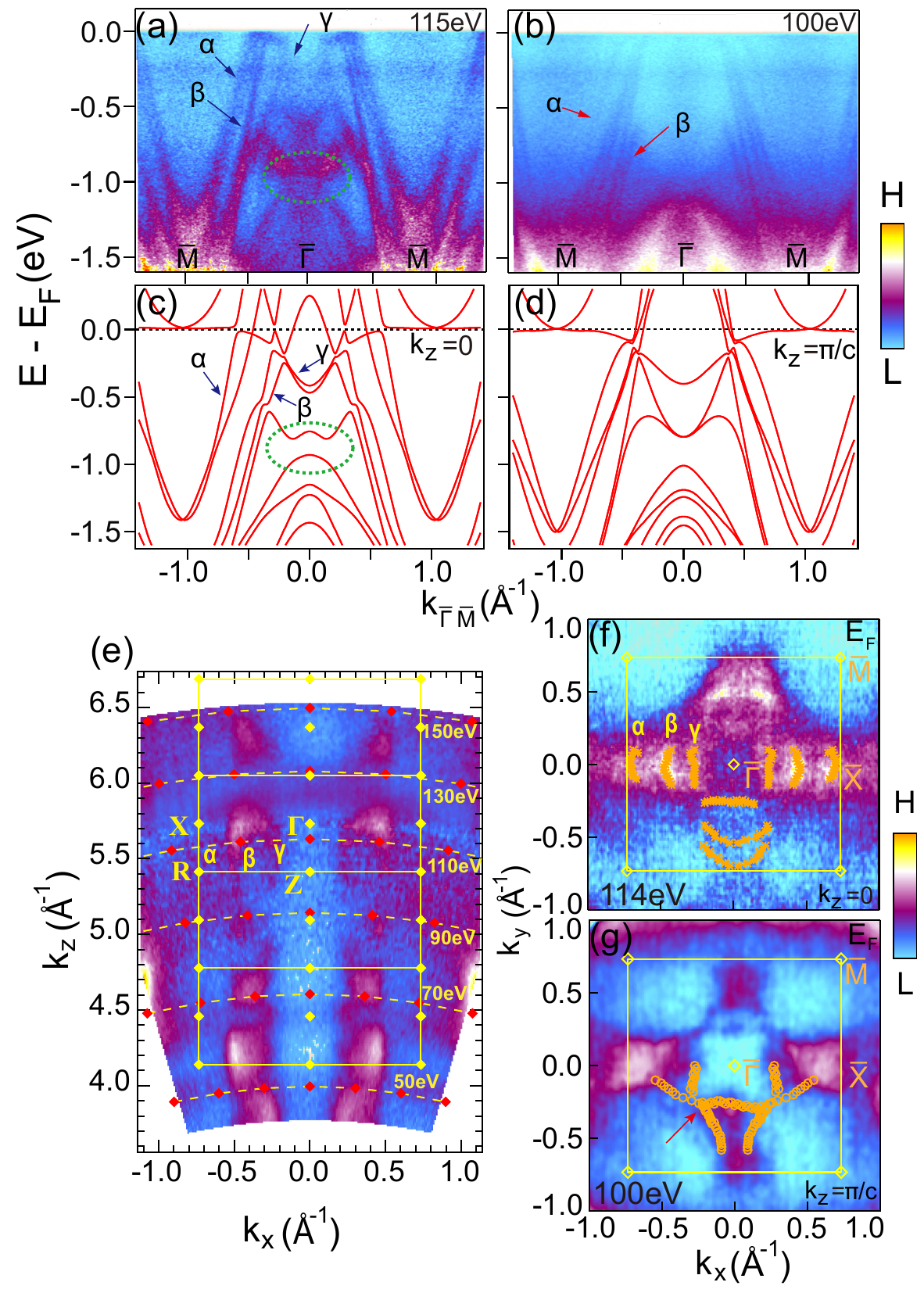}
\caption{(a-d) ARPES spectra taken with 115 eV (a) and 100 eV (b) photons along the $\bar{\Gamma}-\bar{M}$ direction, in comparison with DFT calculations at $k_z$ = 0 (c) and $\pi/c$ (d). \textcolor{red}{(e)} The $k_x$-$k_z$ map at $E_{F}$ along the in-plane $\bar{\Gamma}-\bar{X}$ direction ($k_x$) using vertically polarized photons. The yellow rectangles (with points) mark the BZ boundaries. Red dashed curves represent various $k_z$ cuts corresponding to different photon energies. (f,g) In-plane FS maps from 114 eV (f) and 100 eV (g) photons. The orange crosses and circles in the lower halves of the FS maps indicate experimentally extracted Fermi contours. The color bars for the image plots are indicated on the right (L: low intensity, H: high intensity).
}
\label{fig2}
\end{figure}

\par Figure ~\ref{fig2}(a,b) show the dispersion of conduction bands along $\bar{\Gamma}-\bar{M}$ taken with two representative photon energies (additional data and calculations in Fig. S3-S4 of \cite{supplementary}). Except from the flat $4f$ bands at the Fermi level ($E_F$) and -0.27 eV, which we shall discuss later, the dispersive conduction (non-$4f$) bands well below $E_F$ consist of large electron pockets centered at $\bar{M}$ (denoted as $\alpha$), as well as both hole ($\beta$) and shallow electron ($\gamma$) pockets centered at $\bar{\Gamma}$ (blue arrows in Fig. ~\ref{fig2}(a)). The observed conduction bands away from $E_F$ can be compared reasonably well with the DFT calculations (Fig. ~\ref{fig2}(c,d)), which treat Ce $4f$ electrons as core electrons (core-$4f$ calculation). For example, the $\alpha$ and $\beta$ bands show small differences for $k_z$ = 0 and $k_z$ = $\pi/c$ (compare Fig. ~\ref{fig2}(a,c) with Fig. ~\ref{fig2}(b,d)), and there is a M-shape band near -1 eV at $k_z$ = 0 (green dashed circles in Fig. ~\ref{fig2}(a,c)). The $k_z$ dispersion can be further verified by the photon-energy dependent scan in Fig. ~\ref{fig2}(e), where the band positions at $E_F$ exhibit weak periodic modulation with $k_z$, allowing for an estimation of the inner potential $V_0$ $\approx$ 15 eV. Despite some $k_z$-dependent intensities caused by weak $k_z$ dispersion, the dominant spectral features are nearly vertical and extend along the $k_z$ direction, indicating its quasi-2D nature. The quasi-2D FS implies dominant intralayer hopping compared to the weak interlayer hopping, which is thought to be crucial for the high-field odd-parity superconducting phase in CeRh$_{2}$As$_{2}$.

\par From the DFT calculations, both $\alpha$ and $\beta$ bands actually consist of two closely spaced (or split) bands, which can be clearly identified in the ARPES data (red arrows in Fig. ~\ref{fig2}(b)). Direct observation of these fine features demonstrates the high spectral quality. These splittings are closely related to the CaBe$_{2}$Ge$_{2}$-type crystal structure, which hosts alternating Rh-As-Rh and As-Rh-As trilayers that break the local inversion symmetry of the Ce layers. Indeed, calculations assuming the ThCr$_2$Si$_2$-type structure with identical As-Rh-As trilayers, where the centrosymmetricity of the Ce layers is restored, do not exhibit such band splitting (Fig. S5 in \cite{supplementary}). Therefore, these fine splittings are direct manifestations of the local noncentrosymmetricity of the Ce layers, which is intimately connected to the multiple superconducting phases in CeRh$_{2}$As$_{2}$ \cite{Khim2021science,PhysRevX.AngledependenceHc2,PhysRevB2012PDW,PhysRevResearch.3.L032071,PhysRevResearch.3.023204,PhysRevResearch.3.033133,PhysRevB.105.L020505,PhysRevResearch.3.023179,PhysRevB.104.134517,arXiv2022theoryofLISB}. We note that for the HF compounds with the ThCr$_2$Si$_2$-type centrosymmetric structure, e.g., CeRu$_2$Si$_2$ \cite{denlinger2001comparative} and CeCu$_2$Si$_2$ \cite{Wu2021CeCu2Si2}, no similar band splittings have ever been reported.

\par The in-plane FS maps at $k_z$ = 0 and $\pi/c$ are shown in Fig. ~\ref{fig2}(f,g), where the Fermi contours extracted experimentally (yellow crosses/circles) are overlaid in the lower halves of the experimental data. For $k_z$ = 0 (Fig. ~\ref{fig2}(f)), the $\alpha$ and $\beta$ pockets exhibit diamond shape (more discussions below), while the inner $\gamma$ pocket appears to be square-like. By contrast, at $k_z$ = $\pi/c$ (Fig. ~\ref{fig2}(g)), the $\alpha$ and $\beta$ pockets become very close and lead to one large diamond-shaped pocket (highlighted by a red arrow), joined by strong spectral features near $\bar{X}$.

\begin{figure}[ht]
\includegraphics[width=1.0\linewidth]{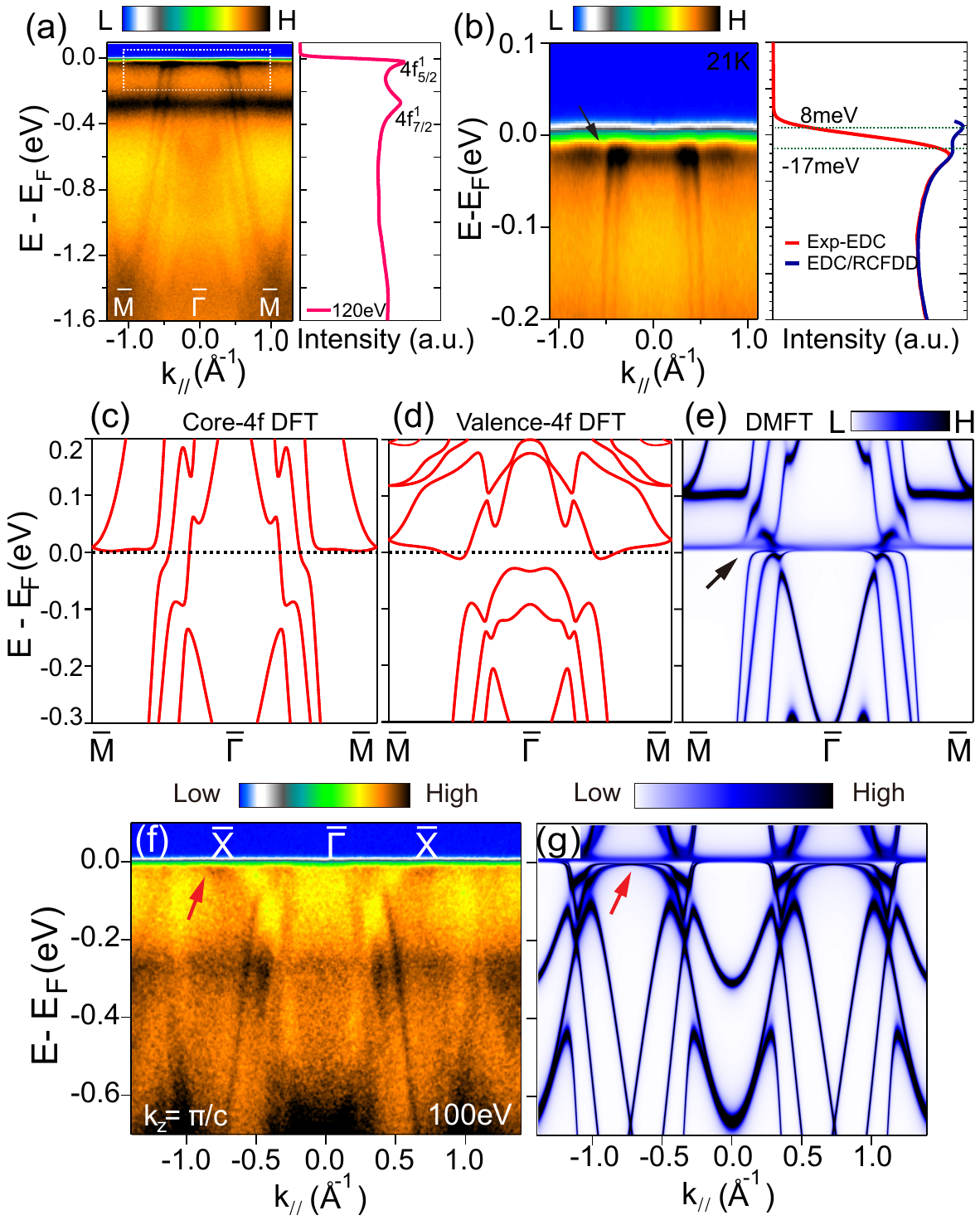}
\caption{(a) Resonant ARPES spectra (left) along $\bar{\Gamma}-\bar{M}$ and its corresponding momentum-integrated EDC (right). (b) Zoom-in view of (a) near $E_{F}$. Right: the EDC (red) and its division by RC-FDD (blue). The EDCs in (a,b) are integrated from -1.2 to 1.2 {\AA}$^{-1}$. (c-e) Band dispersion from core-$4f$ (c), valence-$4f$ (d) DFT calculations and $k$-resolved spectral function from DFT+DMFT calculations (e), for comparison with (b). The electron-like $\gamma$ bands at $\bar{\Gamma}$ are not obvious in experiment likely due to small photoemission matrix elements. (f,g) ARPES spectrum along $\bar{\Gamma}-\bar{X}$ at $k_z$ = $\pi/c$ (f), in comparison with DFT+DMFT calculation (g).
}
\label{fig3}
\end{figure}

\par Now we discuss the $4f$ states near $E_F$. We employed resonant ARPES measurement at 120 eV (the Ce $N$ edge), where the photoemission cross section for the $4f$ states near $E_{F}$ is largely enhanced \cite{sekiyama2000probing}. The results (Fig. ~\ref{fig3}(a)) show that the $4f$ states near $E_F$ actually consist of the $4f^1_{5/2}$ peak near $E_{F}$ ($J=5/2$) and the spin-orbit split $4f^1_{7/2}$ peak at $\sim$-0.27 eV ($J=7/2$), which can be clearly seen in the energy distribution curves (EDCs). These Kondo peaks arise from the many-body Kondo process and develop into propagating heavy bands in a periodic lattice \cite{denlinger2001comparative,jw2005kondo,fujimori2016band}. A zoomed-in view of the quasiparticle dispersion near $E_F$, as shown in Fig. ~\ref{fig3}(b), reveals a nearly flat $4f^1_{5/2}$ band and a clear bending of the conduction bands (black arrow), due to hybridization between conduction and $4f$ electrons ($c$-$f$ hybridization) \cite{Im2008,PhysRevB.2019CeCoIn5,Jang2020,kirchner2020colloquium,PRL164YW,Yuan2021Angle}. To compare with experiments, we perform both DFT and DFT+DMFT calculations (Fig. ~\ref{fig3}(c-e)). Although the core-$4f$ DFT calculation (Fig. ~\ref{fig3}(c)) can capture the conduction bands away from $E_F$, it cannot account for the correlated $4f$ bands near $E_F$. By contrast, the valence-$4f$ calculation (Fig. ~\ref{fig3}(d)), which treats $4f$ electrons as valence electrons with Hubbard $U$, can generate $4f$ bands near $E_F$, but their position and dispersion is very different from experiment - such deviation for the $4f$ bands is not surprising in strongly correlated HF systems. Fig. ~\ref{fig3}(e) shows the calculated $k$-resolved spectral function from DFT+DMFT, which reveals very good agreement with experiments: the weakly dispersive $4f$ band and the bending of conduction bands near $E_F$ can be reproduced very well. The bending of conduction bands (black arrows in Fig. ~\ref{fig3}(b,e)), caused by the $c$-$f$ hybridization, is also accompanied by an increase of $4f$ weight, leading to characteristic "hot" spots near $E_F$ (Fig. ~\ref{fig3}(b)). Figure ~\ref{fig3}(f,g) demonstrate the comparison between ARPES and DFT+DMFT along $\bar{\Gamma}-\bar{X}$ at $k_z$ = $\pi/c$, where excellent agreement can be found. In particular, the heavy hole band near $\bar{X}$ (red arrows in Figure ~\ref{fig3}(f,g)) gives rise to a van Hove singularity (VHS) at $E_F$ with large density of states (DOS). This VHS, different from the conduction-band VHS at $k_z$ = 0 reported in an independent ARPES paper \cite{ChenXuezhi2023arXiv}, is driven by $c$-$f$ hybridization characteristic of HF systems (Fig. S3 in \cite{supplementary}).

\begin{figure}[ht]
\includegraphics[width=1.0\columnwidth]{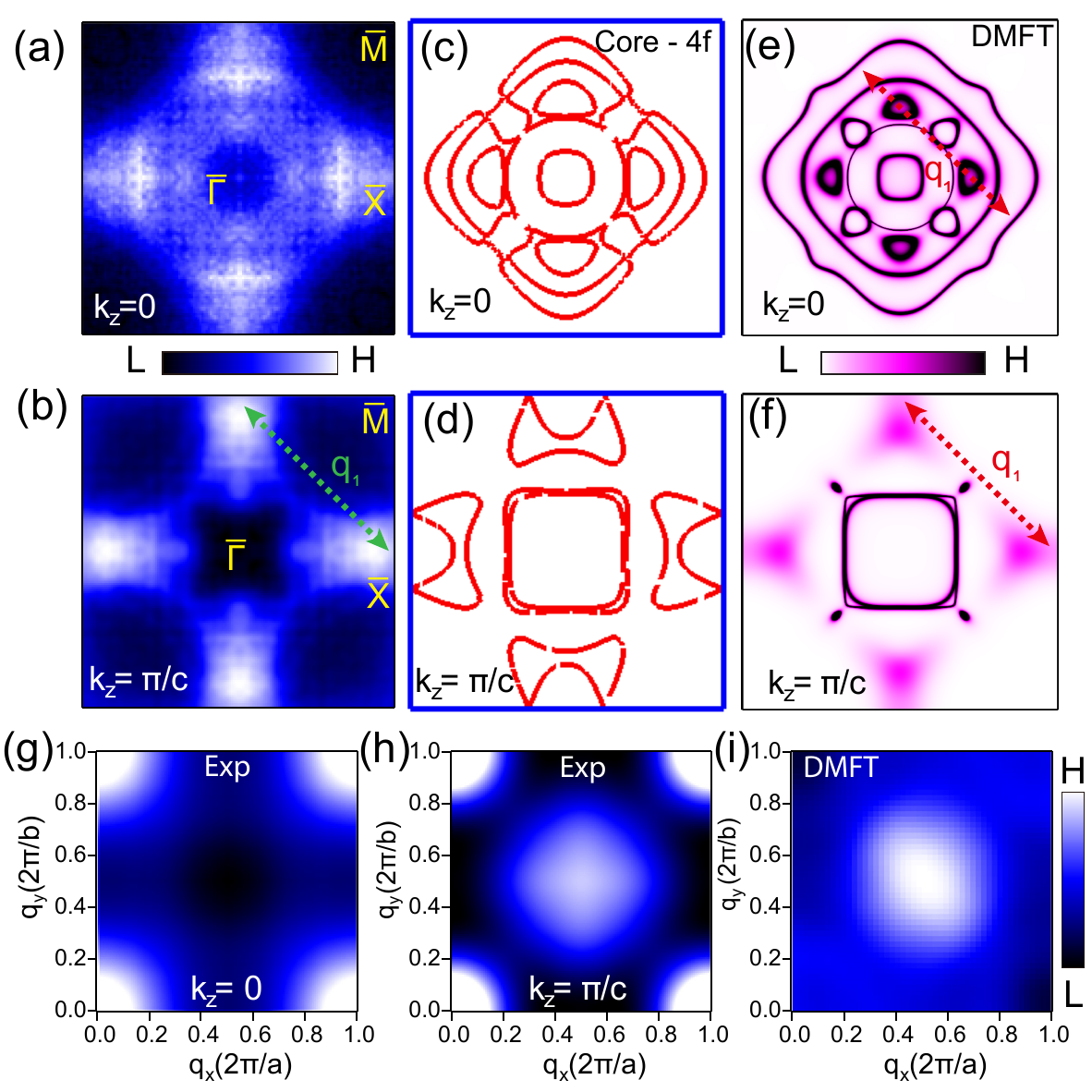}
\caption{ (a,b) In-plane experimental FS maps within first BZ at $k_z$ = 0 (a) and $\pi/c$ (b). Here the maps are symmetrized according to the bulk $C_4$ rotation. (c-f) The calculated in-plane FS's from the core-$4f$ DFT (c,d) and DFT+DMFT (e,f) calculations. The dashed arrows in (b,e,f) indicate the nesting vector $q_1$. (g,h) Autocorrelation $C(q,0)$ of the ARPES FS maps for $k_z$ = 0 (g) and $\pi/c$ (h). (i) The quadrupolar susceptibility $\chi(q)$ calculated from DFT+DMFT.
}
\label {fig4}
\end{figure}

\par The unusual QDW at $T_{0}$ in CeRh$_{2}$As$_{2}$ was attributed to mixing of the ground-state and first excited $4f$ doublets due to strong Kondo effect \cite{PhysRevX.12.011023}, which effectively leads to a quartet ground state for $4f$ electrons, similar to cubic systems with quadrupole order, e.g., CeB$_{6}$ \cite{ASC2016multipolar,Koitzsch2016Nesting,Jang2017Large}. Since the crystal electric field (CEF) states of $4f$ electrons can manifest through the fine satellite structures in the Kondo peaks \cite{PhysRevB.2019CeCoIn5,PRB2018CeIrIn5,ehm2007high,patil2016arpes}, we divided the ARPES spectra by the resolution-convoluted Fermi-Dirac distribution (RC-FDD) (Fig. S6 in \cite{supplementary}), which recovers the full spectral function near $E_{F}$. The result (right panel in Fig.~\ref{fig3}(b)) reveals two $4f$ peaks at $\sim$8 meV and $\sim$-17 meV, respectively. Previous studies suggest that the energy separation between the ground state $\Gamma_7^{1}$ and first excited doublet $\Gamma_6$ (second excited doublet $\Gamma_7^{2}$) is $\Delta_1$ $\sim$ 2.6 meV ($\Delta_2$ $\sim$ 15.5 meV), respectively \cite{Khim2021science,PhysRevX.12.011023,Christovam2023arXiv}. Therefore, the satellite peak at -17 meV could correspond to CEF excitations related to $\Gamma_7^{2}$, while the peak at $\sim$8 meV might contain broadened contributions from both $\Gamma_7^{1}$ and $\Gamma_6$, due to limited energy resolution. The observed fine structures of the $4f^1_{5/2}$ Kondo peak, as well as their temperature evolution (Fig. S7 in \cite{supplementary}), are overall consistent with the proposed CEF scheme \cite{Khim2021science,PhysRevX.12.011023,Christovam2023arXiv}.

\par Figure ~\ref{fig4}(a-f) show the detailed comparison between the experimental and theoretical FS's. While the core-$4f$ DFT calculation appears to explain the shape of the experimental FS at $k_z$ = 0 (compare Fig. ~\ref{fig4}(a,c)), it fails to account for the FS at $k_z$ = $\pi/c$ (compare Fig. ~\ref{fig4}(b,d)). By contrast, the DFT+DMFT calculations in Fig. ~\ref{fig4}(e,f) show very good agreement with experiments at both $k_z$ cuts, particularly the large diamond-shaped pocket and the strong spectral feature near $\bar{X}$ at $k_z$ = $\pi/c$ (compare Fig. ~\ref{fig4}(b,f)). Note that the DFT+DMFT FS at $k_z$ = 0 has similar shape as the core-$4f$ FS (compare Fig. ~\ref{fig3}(c,e)), although the pocket sizes are slightly smaller due to band bending from $c$-$f$ hybridization (Fig. ~\ref{fig3}(b,e)). Therefore, the experimental FS and its anisotropic $4f$ weight is a direct consequence of the band-dependent $c$-$f$ hybridization, which is well captured by DFT+DMFT calculations.

\par Both the experimental and theoretical FS maps exhibit signatures of nesting. A nesting vector $q_1$ $\approx$ ($\pi/a$, $\pi/a$) marked in Fig. ~\ref{fig4}(b,f) can connect the heavy quasiparticles with high DOS at $\bar{X}$ (see also Fig. ~\ref{fig3}(f,g)). This vector can also link the parallel portions of the outer $\alpha$ pocket at $k_z$ = 0, as indicated by a red dashed arrow in Fig. ~\ref{fig4}(e). For quantitative analysis of FS nesting, we calculate the autocorrelation $C(q,\omega)$ of the ARPES spectral function $A(\mathbf{k},\omega)$ via \cite{PhysRevLett.96.067005,PhysRevLett.96.107006,PhysRevLett.99.216404}
$$C(q,\omega) = \sum_{\mathbf{k}} A(\mathbf{k},\omega) A(\mathbf{k}+q,\omega)$$,
where the summation is over the first BZ. The 2D results for the $k_z$ = 0 and $k_z$ = $\pi/c$ FS maps are shown in Fig. ~\ref{fig4}(g,h), respectively. A pronounced peak at $q$ = ($\pi/a$, $\pi/a$) in Fig. ~\ref{fig4}(h) indicates FS nesting at this wave vector. Note that there is no obvious peak at ($\pi/a$, $\pi/a$) in the $k_z$ = 0 map in Fig. ~\ref{fig4}(g). This is due to the weak photoemission intensity of the outer $\alpha$ pocket along the diagonal $\bar{\Gamma}-\bar{M}$ direction (see Fig. ~\ref{fig4}(a)), where large contributions to the FS nesting are expected (red dashed arrow in Fig. ~\ref{fig4}(e)). The FS nesting can be further verified by the DFT+DMFT calculation in Fig. ~\ref{fig4}(i), where the quadrupolar susceptibility $\chi(q)$ using the three-dimensional electronic structure obtained from the DFT+DMFT calculation is plotted \cite{supplementary}. The result shows a strong peak at $q$ = ($\pi/a$, $\pi/a$), confirming the FS instability associated with this wave vector.

\par In HF systems, the nesting of FS with heavy quasiparticles is often necessary for the strong magnetic excitations and spin fluctuations \cite{Eremin2008Feedback,2011NPStockert,Smidman2023RMP}, which could eventually lead to HF superconductivity. At the nesting vector (usually the same as the AFM ordering vector in parent compound), spin resonances have been detected \cite{2011NPStockert,Sato2001Strong,Stockert2023SR}, which are closely connected to the superconducting order parameter. Interestingly, the observed FS nesting at $q_1$ = ($\pi/a$, $\pi/a$) in CeRh$_{2}$As$_{2}$ matches well with the quasi-2D AFM excitations at ($\pi/a$, $\pi/a$) from INS measurements \cite{Chen2023arXiv}. It is therefore desirable to check if spin resonances can be observed at this vector upon entering the superconducting state.

\par In principle, the proposed QDW at $T_{0}$ also requires nesting of the FS \cite{PhysRevX.12.011023}. The coexistence of strong magnetic excitations and quadrupole order in CeRh$_{2}$As$_{2}$ is reminiscent of the famous compound URu$_{2}$Si$_{2}$ \cite{Mydosh2011Hidden}, where a "hidden-order" transition at $\sim$17.5 K could be attributed to certain type of multipole order \cite{Ikeda2012Emergent}, accompanied by strong magnetic excitations observed by INS \cite{Broholm1987Magnetic,Broholm1991Magnetic,Wiebe2007Gapped}. In addition, extensive ARPES studies of URu$_{2}$Si$_{2}$ revealed partial gapping of the nested FS across the "hidden-order" transition \cite{Santander2009Fermi,Boariu2013Momentum,Chatterjee2013PRL,Meng2013Imaging,Yoshida2010Signature,Zhang2018ARPES,Denlinger2022Global}, although the underlying mechanism remains unsettled. Therefore, CeRh$_{2}$As$_{2}$ could provide another interesting platform to study the delicate interplay between orbital and magnetic degrees of freedom in correlated $f$-electron systems, as well as their connection with the unconventional superconductivity \cite{PhysRevB.106.L140502}.

\par According to theory, a large Rashba-type SOC due to local noncentrosymmmetricity and a small interlayer hopping are both crucial for realizing the field-induced parity transition in CeRh$_{2}$As$_{2}$ \cite{PhysRevB2012PDW,PhysRevResearch.3.L032071,PhysRevResearch.3.023204,PhysRevResearch.3.033133,PhysRevB.105.L020505}. Our observation of fine band splittings caused by the locally noncentrosymmetric structure (Fig. ~\ref{fig2}(a,b)) and a quasi-2D FS with strong $4f$ contributions (Fig. ~\ref{fig2}(e) and Fig. S8 in \cite{supplementary}) provides direct spectroscopic proof of these key factors. The quantitative information obtained herein lays the basis for more in-depth investigations in the future. Our results may also help explain why the superconducting parity transition, originally proposed in the ideal 2D systems \cite{PhysRevB2012PDW}, can be realized in bulk CeRh$_{2}$As$_{2}$ \cite{Khim2021science} - since the superconductivity is mainly driven by the quasi-2D $4f$ bands from the Ce layers.

\par In summary, our high-resolution ARPES measurements on CeRh$_{2}$As$_{2}$, in combination with DFT+DMFT calculations, demonstrate that the Ce $4f$ electrons play a significant role in the FS due to band-dependent $c$-$f$ hybridization. We find that the FS exhibits clear nesting at ($\pi/a$, $\pi/a$) and the nesting is facilitated by a VHS near $\bar{X}$ at $k_z$ = $\pi/c$, originating from $c$-$f$ hybridization. The FS nesting can well explain the observed AFM excitations at ($\pi/a$, $\pi/a$) from INS measurements \cite{Chen2023arXiv}, which might be the driving force for its superconductivity. Our results further unveil the fine band splittings related to the locally noncentrosymmetric structure, a quasi-2D FS due to weak interlayer hopping and strong Kondo peaks likely caused by low-lying CEF excitations. The spectroscopic insight can be important to understand the electronic and magnetic excitations in CeRh$_{2}$As$_{2}$, as well as the origin of multiple superconducting phases in this enigmatic compound.

\par This work is supported by the National Key R$\&$D Program of China (Grant No. 2022YFA140220, No. 2023YFA1406303), the State Key project of Zhejiang Province (No. LZ22A040007), the National Science Foundation of China (No. 12174331, 12204159, 12274364), the Key R$\&$D Program of Zhejiang Province, China (2021C01002), and the Bridging Grant (BG 11-072020) with China, Japan, South Korea and ASEAN region funded by the Swiss State Secretariat for Education, Research and Innovation. We acknowledge MAX IV Laboratory for time on Beamline BLOCH under Proposal 20200306. Research conducted at MAX IV, a Swedish national user facility, is supported by the Swedish Research council under contract 2018-07152, the Swedish Governmental Agency for Innovation Systems under contract 2018-04969, and Formas under contract 2019-02496. We thank Dr. Craig Polley, Ms. Dongting Zhang, Mr. Yuxin Chen, Dr. Tong Chen, Prof. Yu Song, Prof. Michael Smidman and Prof. Xin Lu for experimental assistance or helpful discussion.

\par Note: during the submission process of our paper, we became aware of an independent ARPES work on CeRh$_{2}$As$_{2}$ \cite{ChenXuezhi2023arXiv}, which emphasizes the coexistence of the VHS near $X$ at $k_z$ = 0 and the $4f$ flat bands.

%

\end{document}